\documentclass[prl,twocolumn,superscriptaddress]{revtex4-1}
\usepackage{txfonts}
\usepackage{bm}
\usepackage{graphicx}

\begin{document}
\title{Excitation of the Higgs Mode in a Superfluid Fermi Gas in the BCS-BEC Crossover}

\author{Jun Tokimoto}
\email{1214705@ed.tus.ac.jp}
\affiliation{Department of Physics, Tokyo University of Science, 1-3 Kagurazaka, Shinjuku-ku, Tokyo, 162-8601, Japan}

\author{Shunji Tsuchiya}
\affiliation{Department of Physics, Chuo University, 1-13-27 Kasuga, Bunkyo-ku, Tokyo 112-8551, Japan}

\author{Tetsuro Nikuni}
\affiliation{Department of Physics, Tokyo University of Science, 1-3 Kagurazaka, Shinjuku-ku, Tokyo, 162-8601, Japan}

\begin{abstract}
In quantum many-body systems with spontaneous breaking of  continuous symmetries, Higgs modes emerge as collective amplitude oscillations of order parameters. Recently, Higgs mode has been observed in the ultracold Fermi gas. In the present paper, we use the time-dependent Bogoliubov-de Gennes equations to investigate Higgs amplitude oscillations of the superfluid order parameter in a Fermi gas induced by a rapid change of the ${\it s}$-wave scattering length. In particular, we investigate the Higgs mode with different values of the initial scattering length. We find that the energy of the Higgs mode coincides with the threshold energy of the pair-breaking excitation, and exponent of the power-low decay of the Higgs mode $\gamma$ continuously changes between $\gamma=-1/2$ and $\gamma=-3/2$ through the Bardeen-Cooper-Schrieffer-Bose-Einstein condensation (BCS-BEC) crossover. Moreover, we propose the optimal ramp speed of the scattering length for observing the clearest Higgs oscillations.
\end{abstract}
\maketitle
Higgs modes are collective modes associated with amplitude fluctuations of order parameters in systems involving spontaneous breaking of continuous symmetries \cite{higgs-64,sooryakumar-81,littlewood-81,varma-02,pekker-15}. Triggered by recent observations of Higgs modes in various systems including superconductors \cite{matsunaga-13,matsunaga-14,measson-14,sherman-15,katsumi-18}, quantum spin systems \cite{ruegg-08,jain-17,hong-17,souliou-17}, charge-density-wave materials \cite{demsar-99,yusupov-10}, and ultracold atomic gases \cite{bissbort-11,endres-12,leonard-17,behrle-18}, there has been growing interest in Higgs modes in condensed matter systems. In particular, Higgs modes have recently been observed in a ultracold Fermi gas by inducing a periodic modulation of the amplitude of the superfluid gap function $\Delta$ \cite{behrle-18}; this system allows us to study evolution of the Higgs mode in a fermionic superfluid across the BCS-BEC crossover  \cite{Yuzubashyan-05,hannibal-15,bruun-14} and provides with an ideal playground to simulate Higgs modes in superconductors and, moreover, the Higgs particle in the standard model of particle physics \cite{higgs-64} due to its high controllability in experiments.  
\par
Small amplitude oscillations of the order parameter $\Delta$ in a uniform ${\it s}$-wave Fermi superfluid were investigated theoretically by Volkov and Kogan~\cite{volkov-73} in the BCS regime and Gurarie~\cite{gurarie-09} in the BEC regime. In both regimes, the energy of the Higgs mode coincides with the threshold for the creation of fermionic excitation by pair-breaking $2\Delta_{\rm gap}$, where $\Delta_{\rm gap}$ is the gap of fermionic excitations. The threshold behavior of the response function associated with the fluctuation of the order parameter leads to non-exponential damping of the Higgs mode; it has been predicted that modulation of the amplitude of the gap function exhibits power-law decay in time, where the decay exponents are predicted to be $\gamma=-1/2$ in the BCS regime \cite{volkov-73} and $\gamma=-3/2$ in the BEC regime \cite{gurarie-09}. Furthermore, Gurarie predicted the abrupt change of the power at the $\mu=0$~\cite{gurarie-09}. More recently, Scott et al.~\cite{scott-12} used Bogoliubov-de Gennnes (BdG) equations to study the dynamics of the order parameter $\Delta$ in a superfluid Fermi gas, and confirmed the $\vert \Delta \vert$ exhibits oscillations with the frequency $\Delta_{\rm gap}/\hbar$ in both BCS and BEC regimes. However, detailed and quantitative analysis of the frequency and decay exponent of the Higgs mode in the entire regimes in the BCS-BEC crossover is still lacking.
\par
In this paper, we study the Higgs mode in a superfluid Fermi gas; we numerically simulate modulation of the amplitude of the gap function in time in a uniform system as well as in a trapped system. Specifically, we calculate the frequency and decay exponent of amplitude modulation through the BCS-BEC crossover to compare with the prediction in Ref.~\cite{gurarie-09}. Additionally, we propose the optimal ramp speed of the change of {\it s}-wave scattering length for observing the clearest oscillations in experiments.
\par
 
We study time-evolution of a superfluid gap $\Delta(\bm r,t)$ using the time-dependent Bogoliubov-de Gennes equations within the mean-filed level  \cite{scott-12}:
\begin{equation}
 i\hbar\frac{\partial}{\partial t}
  \left(
   \begin{array}{ccc} 
   u_{\nu}(\bm r ,t)\\
   	v_{\nu}(\bm r ,t)
   \end{array}
  \right)=\left(
   \begin{array}{ccc}
  	\hat{h}& \Delta \\
    	\Delta^{*} & -\hat{h}
   \end{array}
  \right)
   \left(
    \begin{array}{ccc}
    	u_{\nu}(\bm r ,t) \\
    	v_{\nu}(\bm r ,t)
    \end{array}
  \right)
  \label{eq.tdBdG}
  \end{equation}
where  $\hat{h}=\hbar^2\nabla^2/2m+U(\bm r)-\mu$, $m$ is the atomic mass, $\mu$ is the chemical potential and $U(\bm r)$ is the trapping potential. These equations have to be solved self-consistently in every time step together with the gap equation 
\begin{equation}
\Delta(\bm r,t)=-g\sum_{\nu}u_{\nu}(\bm r,t)v^{*}_{\nu}(\bm r,t), \label{eq.gap}
\end{equation}
and the number equation
\begin{equation}
	N=2\int d\bm r\sum_{\nu}\vert u_{\nu} (\bm r,t)\vert^2, \label{eq.number}
\end{equation}
where $N$ is the total number of particles.
The coupling constant $g(>0)$ is related with the parameter $1/k_{{\rm F}}a$ via $1/k_{{\rm F}}a= 8 \pi E_{{\rm F}}/(g k_{{\rm F}}^{3})+\sqrt{4E_{{\rm c}}/(\pi E_{{\rm F}})}$ \cite{giorgini-08}, where $E_{{\rm F}}$, $k_{{\rm F}}$, $E_{{\rm c}}$ and $a$ are the Fermi energy, Fermi wavenumber, cut-off energy introduced in the sum of the gap equation and ${\it s}$-wave scattering length, respectively.
\par 
\begin{figure}[h]
\centering\includegraphics[clip,width=8.0cm]{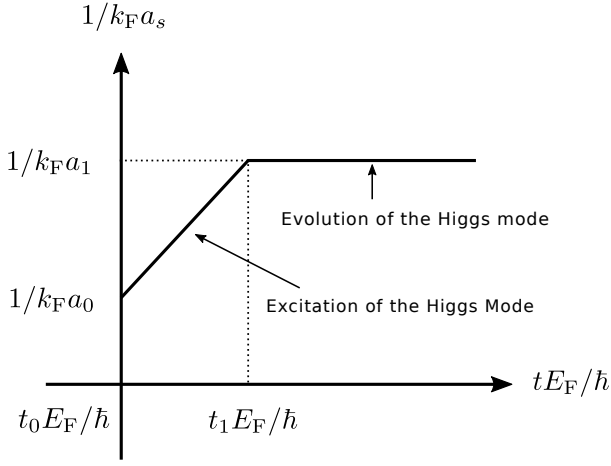} 
\caption{The procedure of our simulation. To excite the Higgs mode, let $1/k_{{\rm F}}a$ increase linearly from $1/k_{{\rm F}}a_{0}$ to $1/k_{{\rm F}}a_{1}$ over the time $t_{1}$. Later on($t>t_{1}$), the $1/k_{{\rm F}}a$ is fixed to $1/k_{{\rm F}}a_{1}$.}
\label{procedure} 
\end{figure}
\par 
We numerically solve the tdBdG equation with splitting method and adapt the fast Fourier transformation (FFT) to the differential term. We excite the Higgs mode following the protocol proposed in Ref.~\cite{scott-12}: modulation of the gap function is induced by rapidly changing the scattering length, which can be realized in experiments by using the Feshbach resonance. Figure~\ref{procedure} shows the protocol of our simulation: starting with the initial state solution at $t=t_0$, the scattering length is decreased so that $1/a k_{{\rm F}}$ is linearly increased from $1/a_{0}k_{{\rm F}}$ to $1/a_{1}k_{{\rm F}}$ over the time interval between $t=t_0$ and $t=t_1$. Then, the scattering length is fixed at $a_{1}$ afterwards ($t>t_1$), as shown in Fig.~\ref{procedure}.
\par
First, we study the uniform Fermi superfluid gas case. For all of the following uniform case results, we consider a periodic boundary conditions with $L=29.18k_{{\rm F}}^{-1}$ in the {\it x}, {\it y} and {\it z} directions, and set the cut-off energy as $E_{c}=15.0E_{\rm F}$ where $E_{\rm F}$ is the Fermi energy. 
Since the system is uniform in the {\it x}, {\it y} and {\it z} directions, the solution of the tdBdG equation can be written as $u_{\nu}({\bm r},t)=u_{{\bm k}}(t)e^{i{\bm k}\cdot{\bm r}}$, $v_{\nu}({\bm r},t)=v_{{\bm k}}(t)e^{i{\bm k}\cdot{\bm r}}$,  where ${\bm k}=(2\pi\alpha_{x}/L_{x}, 2\pi\alpha_{y}/L_{y}, 2\pi\alpha_{z}/L_{z})$ and $\alpha_{x}$,$\alpha_{y}$ and $\alpha_{z}$ are integers.
\par 
\begin{figure}[t]
\begin{center}
\includegraphics[clip,width=8.0cm]{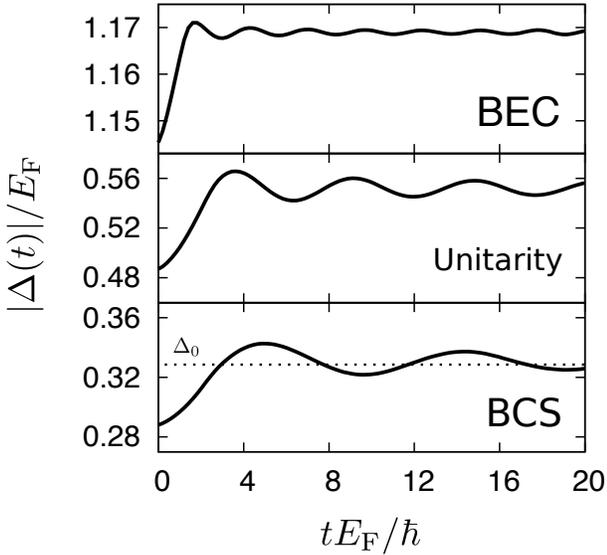}
\caption{The evolution of the order parameter in the BCS ($1/k_{\rm F}a_{0}=-0.6, 1/k_{\rm F}a_{1}=-0.5$), unitary ($1/k_{\rm F}a_{0}=0.1, 1/k_{\rm F}a_{1}=0.2$) and BEC ($1/k_{\rm F}a_{0}=0.6, 1/k_{\rm F}a_{1}=0.7$) regimes.}
\label{higgsexample} 
\end{center}
\end{figure}\par
Figure~\ref{higgsexample} shows  examples of the time evolutions of amplitude of the order parameter $|\Delta(t)|$ for different initial scattering length. 
 We clearly see the oscillations of the amplitude of the order parameter, which we identify with the Higgs mode. 
 One can see the relatively large amplitude of the oscillation in the weak-coupling BCS regime, reflecting the clear nature of the collective mode. With increasing the coupling strength, the oscillation amplitude continuously decreases and thus the collective mode nature becomes less clear in the strong-coupling BEC regime.
\par
To identify the character of these oscillations, we fit the curve in Figs.~\ref{higgsexample} with the following function of time
\begin{equation}\label{fit}
	\vert \Delta(t) \vert = t^{\gamma}A\sin \left( \omega_{\rm H}
						   t + \delta_{\rm H}
						  \right)+\Delta_{\rm 0},
						  \end{equation}
where $\gamma$ is the decay exponent, $\Delta_{\rm 0}$ 
is the equilibrium value of the order parameter, and $\delta_{\rm H}$ is an offset of the phase. We also fitted the curve with the function $\vert \Delta(t) \vert = e^{t\gamma}A\sin \left( \omega_{\rm H}t + \delta_{\rm H}\right)+\Delta_{\rm 0}$ assuming the exponential damping, and found that Eq.\ref{fit} assuming the power-law decay gives the better fitting. 					  
\par
\begin{figure}[t]
\begin{center}
\includegraphics[clip,width=8.0cm]{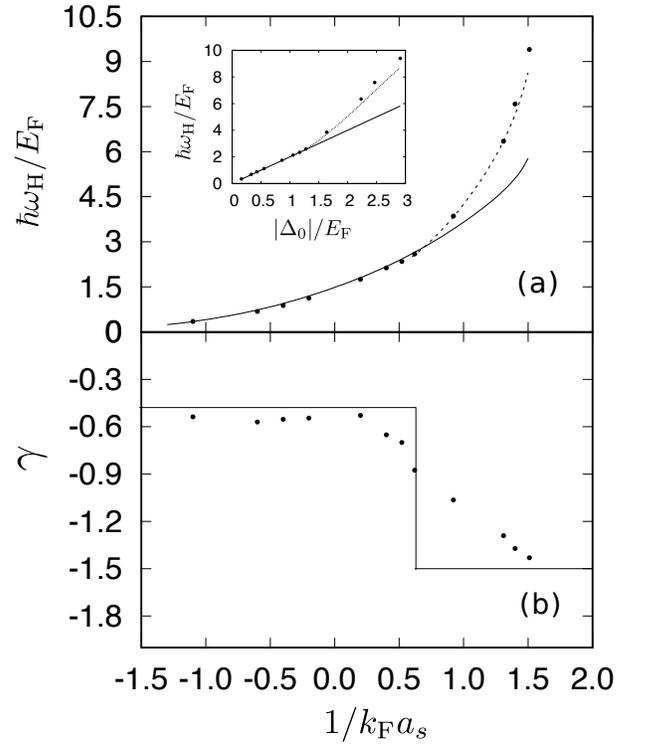}
\caption{(a) The frequency of the Higgs mode in a Fermi superfluid $\hbar \omega_{\rm H}/E_{\rm F}$ against the $1/k_{\rm F}a_{s}$. The solid line indicates $\hbar \omega_{\rm H} = 2|\Delta_{0}|$ and the dashed line indicates $\hbar \omega_{\rm H} = 2\sqrt{\Delta_{0}^2+\mu^2}$. The inset indicates the frequency of the Higgs mode $\hbar \omega_{\rm H}/E_{\rm F}$ against the equilibrium value of the order parameter $\Delta_{0}/E_{\rm F}$. The solid and dashed line indicate the same one as the outer figure. (b) The exponent of the power-low decay $\gamma$ against the $1/k_{\rm F}a_{s}$. The solid line indicate the value predicted by Gurarie in \cite{gurarie-09}.}
\label{homofreqdamp} 
\end{center}  
\end{figure}
Figure~\ref{homofreqdamp}(a) plots the angular frequency of the Higgs mode $\omega_{\rm H}$ obtained from our simulation results, with changing the parameter $1/k_{\rm F}a_{s}$. One can see that the frequency of the Higgs mode $\omega_{\rm H}$ behaves as $\omega_{\rm H}= 2\Delta_{0}/\hbar$ in the BCS side and $\omega_{\rm H}=2\sqrt{\Delta_{0}^2+\mu^{2}}/\hbar$ in the BEC side. In both cases, it can be expressed as $\omega_{\rm H}=2\Delta_{\rm gap}/\hbar$, where $\Delta_{\rm gap}=\sqrt{\Delta_{0}^{2}+\theta(-\mu)\mu^{2}}$ is the gap of the single-particle excitation and $\theta(\mu)$ is the step function. The transition between the BCS and BEC regimes occurs at $\mu=0$.

In the works by Volkov, Kogan~\cite{volkov-73}and Gurarie~\cite{gurarie-09}, it has been shown that the Higgs modes originate from the threshold for creation of the pair-breaking. In the BCS regime, the dominant fermionic excitations are created with finite momenta $p\sim k_{\rm F}$ near the Fermi momentum so that many excitations are collectively involved in the Higgs excitation, while in the BEC regime the dominant fermionic excitations are created near the momentum $p\sim 0$ so that fewer excitations are involved. These predictions are consistent with our simulation results of behaviour of the amplitudes and frequencies of the Higgs oscillations. In particular, we have confirmed that $\omega=2\Delta_{\rm gap}/\hbar$ in the entire regime in the BCS-BEC crossover.  
\par
Figure~\ref{homofreqdamp}(b) shows the decay exponent $\gamma$ in a  superfluid Fermi gas through the BCS-BEC crossover determined from our simulations. One can see that the exponent approaches $\gamma = -1/2$ in the weak coupling BCS limit as Volkov showed in Ref.~\cite{volkov-73} and in the positive $a_{0}$ side, $\gamma$ decreases, and approaches the value $-3/2$, which was predicted by Gurarie \cite{gurarie-09}. In contrast to the prediction by Gurarie \cite{gurarie-09}, in the intermediate (unitarity) regime, our result shows the continuous change of the decay exponent between the value in the BCS regime and BEC regime, rather than the abrupt change at $\mu=0$ predicted by Gurarie~\cite{gurarie-09}. In fact, this continuous behavior of the decay exponent is the expected result within the frame work of the analysis of Refs.~\cite{volkov-73,gurarie-09}, since the rerated response function should smoothly change through the BCS-BEC crossover.

\begin{figure}[t]
\centering\includegraphics[clip,width=8.0cm]{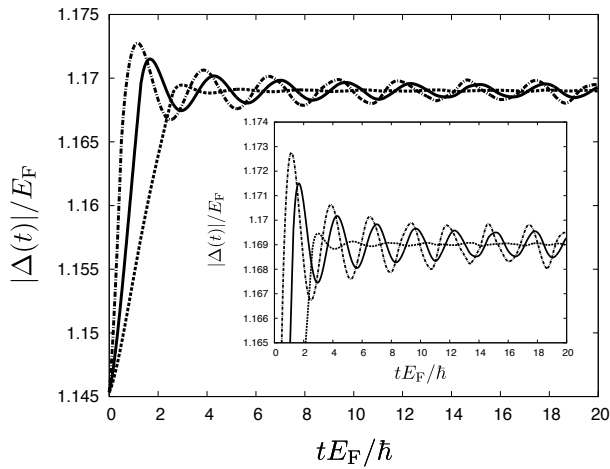} 
\caption{The time evolutions of the order parameter with $t_{1}=0.5\hbar/E_{\rm F}$ (long-dash-solid-dashed), $t_{1}=2.5\hbar/E_{\rm F}$ (dashed) and $t_{1}=1.3\hbar/E_{\rm F}$ (solid). In this case, the half value of the period of the Higgs oscillation $T_{\rm H}/2=1.3\hbar/E_{\rm F}$. The scattering lengthes $1/k_{\rm F}a_{0}=0.4, 1/k_{\rm F}a_{1}=0.45$ are the common parameters in the showed three line.}
\label{clear}
\end{figure}
We now discuss the optimal ramp speed for observing the clearest Higgs mode oscillation. Figure~\ref{clear} compares the Higgs mode oscillations with three different ramp time $t_{1}$. We find that the clearest oscillation is seen for $t_{1}=1.3$, which corresponds to the half the oscillation period of the Higgs mode oscillation. Here we argue that in general, the optimal ramp time $t_{1}$ is given by $t_{1}=T_{\rm H}/2$, where $T_{\rm H}=2\pi/\omega_{\rm H}$ is the oscillation period; it is clear that the ramp speed is fast enough so that one finishes changing the scattering length before the Higgs oscillation occurs. However, if the ramp speed is too first, i.e. $t_{1}<T_{\rm H}/2$, one may excite higher energy modes, leading to noisy oscillations.
We have numerically confirmed that $t_{1}=T_{\rm H}/2$ is the optimal chose for other initial scattering lengths.

Finally we discuss the time evolution of the order parameter of a harmonically trapped superfluid Fermi gas. For all of the following results, we consider a harmonic potential in the {\it x}-direction $U(\bm r)=\frac{1}{2} m\omega_{\rm Trap}^{2}x^{2}$ with trapping frequency $\omega_{\rm Trap}=0.185E_{\rm F}/\hbar$. In {\it y}- and {\it z}-directions, we consider a periodic boundary condition $L_{y}=L_{z}=29.18k_{\rm F}^{-1}$. In this case, the order parameter and density of particles depend on space. Thus the rapid change of the scattering length induces the change both in the order parameter and the density profile, and thus simultaneously excites the Higgs mode and the breathing mode: the fast oscillation with frequency $\omega_{\rm H}$ corresponding to the Higgs oscillation and the slow oscillation with frequency $\omega_{\rm B}$ corresponding to the breathing oscillation. Therefore, we fit the curves of time evolution of the order parameter and the density at the center of the trap with the following functions \cite{tokimoto-17}.
\begin{equation}
\vert \Delta(0,t) \vert = t^{\gamma} A \sin (\omega_{\rm H}t+\delta_{\rm H})+B\sin (\omega_{\rm B}t+\delta_{\rm B})+\Delta_{0}
\end{equation}
\begin{equation}
n(0,t) = C\sin (\omega_{\rm B}t+\delta_{\rm B})+n_{0}
\end{equation}\label{density}
 \begin{figure}[t]
\begin{center}
\includegraphics[clip,width=8.0cm]{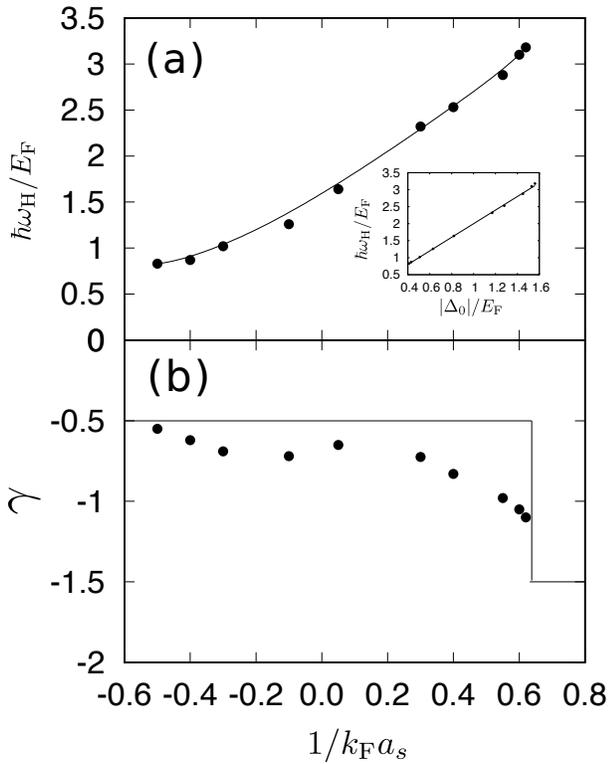}
\caption{(a) The energy of the Higgs mode in a harmonically trapped Fermi superfluid $\hbar\omega_{\rm H}/E_{\rm F}$ against the $1/k_{\rm F}a_{s}$. The inset indicates $\hbar\omega_{\rm F}/E_{\rm F}$ against the equilibrium value of the order parameter at the center of the trap $\Delta_{0}/E_{\rm F}$. The solid line indicates $\hbar \omega_{\rm H} = 2|\Delta_{0}|$ .
(b) The exponent of the power-low decay against the $1/k_{\rm F}a_{s}$. The solid line indicate the value predicted by Gurarie in Ref.~\cite{gurarie-09}.}
\label{higgsharmonicfreq} 
\end{center} 
\end{figure}
In principle, the density fluctuation~\ref{density} may also involve the oscillation associated with the Higgs mode with the frequency $\omega_{\rm H}$. However, our simulation results show that the contribution to the density oscillation from the Higgs mode is negligibly small.
Figure\ref{higgsharmonicfreq}(a) shows the frequency of the Higgs mode in a harmonically trapped Fermi superfluid $\omega_{\rm H}$ against the $1/k_{\rm F}a_{\it s}$.  We emphasize that the entire order parameter collectively oscillates with angular frequency $\omega_{\rm H}$. Figure~\ref{higgsharmonicfreq}(b) shows the decay exponent $\gamma$ in a harmonically trapped Fermi superfluid mainly around the unitarity regime. One can see that our results approach $\gamma=-1/2$ in the weak coupling BCS limit and $\gamma=-3/2$ in the strong coupling BEC limit. In the intermediate regime, one can see the continuous change between $\gamma=-1/2$ and $\gamma=-3/2$. As a whole, Higgs modes in a trapped superfluid behaves in a similar way to a uniform superfluid. 

In conclusion, we have investigated the Higgs mode in a uniform Fermi superfluid gas and harmonically trapped Fermi superfluid gas through the BCS-BEC crossover by solving the time-dependent Bogoliubov-de Gennes equations. We numerically simulated the dynamics of the Higgs amplitude oscillations of the superfluid order parameter induced by a rapid change of the ${\it s}$-wave scattering length. We confirmed that the frequency of the Higgs mode in a Fermi gas coincides the threshold energy of the pair-breaking ($\omega_{\rm H}=2\sqrt{\Delta_{\rm 0}^2+\theta\left( -\mu \right)\mu^2}/\hbar$)  through the BCS-BEC crossover. We also found that the decay exponent $\gamma$ continuously changes from $\gamma=-1/2$ in the weak coupling BCS regime to $\gamma=-3/2$ in the strong coupling BEC regime as $1/k_{\rm F}a$ increases. Moreover, we found that to see the clearest Higgs oscillation one should set the ramp time to half the period of the Higgs oscillation ($t_{1}=T_{\rm H}/2$). The behavior of the Higgs mode in a trapped Fermi superfluid is essentially the same as in a uniform superfluid. As a whole, we conclude that the property of the Higgs mode in a Fermi superfluid continuously change through the BCS-BEC crossover. 
\begin{acknowledgements}
	The authors are grateful to K. Gao for many discussions. In this research work we used the supercomputer of ACCMS, Kyoto University. T.N. was supported by JSPS KAKENHI Grant No. JP16K05504.
\end{acknowledgements}

\end{document}